\documentclass[reprint,aps,prb]{revtex4-1}
\usepackage{graphicx}
\usepackage[utf8]{inputenc}
\usepackage{amsmath}
\usepackage{bm}
\usepackage[caption=false]{subfig}
\usepackage{amsmath}
\graphicspath{{./images/}}
\usepackage{siunitx}

\begin{document}
\title{Tunneling magnetoresistance of perpendicular CoFeB-based junctions \\  with exchange bias}
\author{Orestis Manos$^1$, Alexander B\"ohnke$^1$, Panagiota Bougiatioti$^1$, Robin Klett$^1$, Karsten Rott$^1$, Alessia Niesen$^1$, Jan-Michael Schmalhorst$^1$ and G\"unter~Reiss$^1$\email{Electronic mail: omanos@physik.uni-bielefeld.de}}

\affiliation{$^1$\mbox{Center for Spinelectronic Materials and Devices, Department of Physics,}\\
\mbox{Bielefeld University, Universit\"atsstra\ss e 25, 33615 Bielefeld, Germany}\\}
\date{\today}

\keywords{}

\begin{abstract}
 Recently, magnetic tunnel junctions with perpendicular magnetized electrodes combined with exchange bias films have attracted large interest. In this paper we examine the tunnel magnetoresistance of Ta/Pd/IrMn/Co-Fe/Ta/Co-Fe-B/MgO/Co-Fe-B/capping/Pd magnetic tunnel junctions in dependence on the capping layer, i.e., Hf or Ta. In these stacks perpendicular exchange bias fields of -500\,Oe along with perpendicular magnetic anisotropy are combined. A tunnel magnetoresistance of $(47.2\pm 1.4)\%$ for the Hf-capped sample was determined compared to the Ta one $(42.6\pm 0.7)\%$ at room temperature. Interestingly, this observation is correlated to the higher boron absorption of Hf compared to Ta which prevents the suppression of $\Delta_{\textrm{1}}$ channel and leads to higher tunnel magnetoresistance values. Furthermore, the temperature dependent coercivities of the soft electrodes of both samples are mainly described by the Stoner-Wohlfarth model including thermal fluctuations. Slight deviations at low temperatures can be attributed to a torque on the soft electrode that is generated by the pinned magnetic layer system.
 
\end{abstract}
\maketitle

\section{Introduction}
  Magnetic tunnel junctions (MTJs) are the backbone of modern spintronics. MTJs with a fully epitaxial (001) MgO barrier sandwiched by (001) bcc ferromagnetic electrodes, such as Fe, Co, and CoFe, were first theoretically predicted to show high tunnel magnetoresistance (TMR) of several 100\,$\%$ as a consequence of the coherent tunneling of $\Delta_{\textrm{1}}$ electrons \cite{Butler:2011,Mathon:2001,Zhang:2004}. The experimentally discovered large TMR amplitude of in-plane magnetized MTJs with a crystalline MgO barrier rendered a major breakthrough for these materials \cite{Parkin:2004,Yuasa:2004}.\\
  \mbox{ } Nevertheless, for memory applications, the interest rapidly changes towards out-of-plane magnetized systems. MTJs with  perpendicular magnetic anisotropy (PMA) have several advantages as compared with their in-plane counterparts. Firstly, an increasing density of memory cells on a wafer can be realized since no elliptical shape is required to stabilize the anisotropy direction \cite{Mangin}. Furthermore, the PMA energy is usually much larger than the energy related with the shape anisotropy that can be obtained in planar MTJs, allowing long memory retention at small size \cite{Kishi}. Additionally, for a given retention time, the critical current density to write information by Spin Transfer Torque (STT) switching is strongly reduced, provided that the Gilbert damping remains low enough \cite{Heinonen:2010}. 
  %, magnetic tunnel junctions with perpendicular magnetized electrodes (pMTJs) have been extensively investigated due to their noteworthy properties of high thermal stability and low critical current for spin-transfer-torque induced magnetization switching or precession rendering them as promising candidates for high-density magnetoresistance random access memories[1-2] and nanoscale oscillators[3]. For a plethora of applications pMTJs concentrating the characteristics of large TMR, strong PMA, and robust thermal stability are highly desired.
  However, neighboring MTJs in a memory array as well as the reference layer of the STT-switched MTJ will be magnetically disturbed. This is of major importance since even after a large number of STT switching events the magnetic states of the MTJs do not ``creep" either to some intermediate state or completely reverse. One distinct advantage of MTJs with exchange bias (EB) layers is the robustness of the reference magnetization against such perturbation \cite{Parkin:1999}.\\
%\par
%\vspace{-0.4em}
\mbox{ } Although several investigations on MTJs with EB have been reported \cite{Parkin:2004,Ma,Szulczewski}, the combination of perpendicular MTJs (p-MTJs) with perpendicular exchange bias (PEB) is still challenging. Here, we address this issue by  investigating stacks of p-MTJs with PEB using MnIr/CoFe and CoFeB electrodes and varying the capping of the soft electrode.\\
\mbox{ } There are two primary requirements for p-MTJs with PEB from the magnetic respect. Firstly, the pinned part of the junction must have large PEB along with low coercivity  ($H_{\textrm{c}}$) in order to prevent the simultaneous switching of both electrodes. Secondly, the stacks of the soft and pinned electrode must show high PMA to ensure that the relative orientation of the electrodes' magnetizations can be parallel (low resistance) or antiparallel (high resistance) in the perpendicular direction. Moreover, the bottom part of the junction is preferred for the pinned part, because MnIr acts as an additional seed layer that promotes the (111) texture of the subsequent ferromagnetic layer and therefore higher PMA as van Dijken $et\,al.$ reported \cite{Dijken:2005}.\\
\mbox{ } Although the CoFeB/MgO films directly grown on a MnIr layer exhibit relatively high PEB, one of their major disadvantages is the insufficient PMA. Zhang $et\,al.$  \cite{Zhang:2015} showed that the introduction of an interlayer of CoFe/Ta at the antiferromagnet (AF)/ferromagnet (FM) interface leads to a large PEB with large PMA. Furthermore, the presence of Boron (B) can influence the stack twofold. Firstly, the crystallization of CoFeB is inextricably connected to the B diffusion within the CoFeB layer and to the neighboring layers. In particular, a high B concentration leads to poor crystallization and consequently to small PMA. On the other hand, the presence of B at the interface of CoFeB/MgO is detrimental to TMR \cite{Kodzuka:2012} because it suppresses the conductance through the band of  $\Delta_{\textrm{1}}$ symmetry, which is known to be responsible for high TMR in epitaxial CoFe/MgO/CoFe (001) \cite{Burton:2006}. Therefore, in order to reduce the concentration of B in the CoFeB and at the CoFeB/MgO interface, one option is to insert a B absorber material in close vicinity to the CoFeB\cite{Hindmarch:2011}. We verify in this work that the introduction of material with larger B absorption than Ta (such as Hf) enhances the PMA of the free electrode and leads to larger TMR. %Interestingly, this behavior appears to remain in the whole temperature range between 10 K and 300 K. %Nonetheless, in the abstract the temperature and bias voltage dependence of TMR are in line with the suggested theoretical model of magnon assisted tunneling \cite{Zhang:1997} \cite{Drewello:2008} .  
\vspace{-1em}

\section{Preparation}

The films were deposited on thermally oxidized Si wafers at room temperature (RT) by DC magnetron sputtering, at Ar pressure of 
P = $2\,\cdot\,10^{-3}$\,mbar. The following two types of samples were prepared and investigated
\vspace{0.5em} \\1)\mbox{ \,\,\,\,\,}Ta(4)$\,$/$\,$Pd(2)$\,$/$\,\textrm{Mn}_{\textrm{83}}\textrm{Ir}_{\textrm{17}}$(8)$\,$/$\,\textrm{Co}_{\textrm{50}}\textrm{Fe}_{\textrm{50} }$(1)$\,$/$\,$Ta(0.6)$\,$\ \mbox{ \,\,\,\,\,\,\,\,\,}$\,\textrm{Co}_{\textrm{40}}\textrm{Fe}_{\textrm{40}}\textrm{B}_{\textrm{20}}$(0.8)$\,$/$\,$MgO(2)$\,$/$\,\textrm{Co}_{\textrm{40}}\textrm{Fe}_{\textrm{40}}\textrm{B}_{\textrm{20}}$(1.2)
\mbox{ \,\,\,\,\,\,\,\,\,\,\,}Hf(5)$\,$/$\,$Pd(3) \vspace{0.5em} \\ 2)\mbox{ \,\,\,\,\,}Ta(4)$\,$/$\,$Pd(2)$\,$/$\,\textrm{Mn}_{\textrm{83}}\textrm{Ir}_{\textrm{17}}$(8)$\,$/$\,\textrm{Co}_{\textrm{50}}\textrm{Fe}_{\textrm{50}} $(1)$\,$/$\,$Ta(0.6)$\,$\ 
\mbox{ \,\,\,\,\,\,\,\,\,\,}$\textrm{Co}_{\textrm{40}}\textrm{Fe}_{\textrm{40}}\textrm{B}_{\textrm{20}}$(0.8)$\,$/$\,$MgO(2)$\,$/$\,\textrm{Co}_{\textrm{40}}\textrm{Fe}_{\textrm{40}}\textrm{B}_{\textrm{20}}$(1.2)\\\mbox{ \,\,\,\,\,\,\,\,\,\,\,}Ta(3)$\,$/$\,$Pd(3)
\vspace{0.5em}\\
where the number in parentheses is the nominal thickness of each layer in nm. The layer thicknesses stemmed from an optimization process of a series of films and were chosen for further investigation due to their large PEB and PMA of these stacks. It is known that a strong (111) texture of MnIr leads to enhanced PEB \cite{Chen}. For that reason we used the buffer layer Ta(4)/Pd(2) to induce a strong (111) texture \cite{Zhang:2015}. Ta, Pd, $\textrm{Co}_{\textrm{40}}\textrm{Fe}_{\textrm{40}}\textrm{B}_{\textrm{20}}$, $\textrm{Co}_{\textrm{50}}\textrm{Fe}_{\textrm{50}}$, Hf, $\textrm{Mn}_{\textrm{83}}\textrm{Ir}_{\textrm{17}}$ and MgO films were deposited from %Ta, Pd, Hf, $\textrm{Mn}_{\textrm{83}}\textrm{Ir}_{\textrm{17}$, MgO, $\textrm{Co}_{\textrm{40}}\textrm{Fe}_{\textrm{40}}\textrm{B}_{\textrm{20}}$, $\textrm{Co}_{\textrm{50}}\textrm{Fe}_{\textrm{50} 
elemental and composite targets. The purity of all targets was 99.9$\%$ or higher. All samples were annealed at 280$^{\circ}$C for 60\,min in vacuum ($<3\,\cdot\,10^{-7}$\,mbar) with magnetic field of 7\,kOe applied perpendicular to the film plane, in order to achieve the required coherent (001)-textured bcc crystal structure and induce the PEB.\\
\mbox{ } In the post-annealing procedure the amorphous CoFeB electrodes crystallization starts at the CoFeB/MgO interfaces, templated by the (001) texture in the crystalline MgO tunnel barrier layer\cite{Park:2006,Hindmarch:2011}. Perpendicular hysteresis loops were recorded using the magnetooptical Kerr effect (MOKE). For simplicity, in the rest of the paper the films $\textrm{Co}_{\textrm{40}}\textrm{Fe}_{\textrm{40}}\textrm{B}_{\textrm{20}}$, $\textrm{Co}_{\textrm{50}}\textrm{Fe}_{\textrm{50}}$ and $\textrm{Mn}_{\textrm{83}}\textrm{Ir}_{\textrm{17}}$ will be symbolized as CoFeB, CoFe and MnIr, respectively. On the annealed samples, circular MTJ pillars with diameters of 120$\,$nm, 140$\,$nm, 240$\,$nm and 480$\,$nm were patterned by e-beam lithography and were etched by Ar-ions until only the Ta/Pd layers remained as common bottom lead. 
After etching, 150$\,$nm of Ta$_2$O$_5$ were deposited to insulate the MTJs followed by a lift off procedure. In a subsequent patterning process individual gold contact pads were placed on top of each MTJ and a large gold electrode was placed at the edge of the common bottom contact. All measurements were performed by a conventional two probe technique. Additionally, for the temperature dependent experiments we used a closed cycle helium cryostat by Cryogenic Ltd. in a temperature range of 1.8–300\,K. 
%Normalized hysteresis loops of the individual stacks: (a) MgO(2)/CoFeB(1.2)/Hf(5)/Pd(3), (b) MgO(2)/CoFeB(1.2)/Ta(3)/Pd(3), (c) and (d) Ta(4)/Pd(2)MnIr(8)CoFe(1)Ta(0.6)CoFeB(0.8)MgO(2) acquired via MOKE at RT.

\section{Results and discussion}

\begin{figure}[!ht]
    \centering
    \includegraphics[height=6.7cm, width=\linewidth]{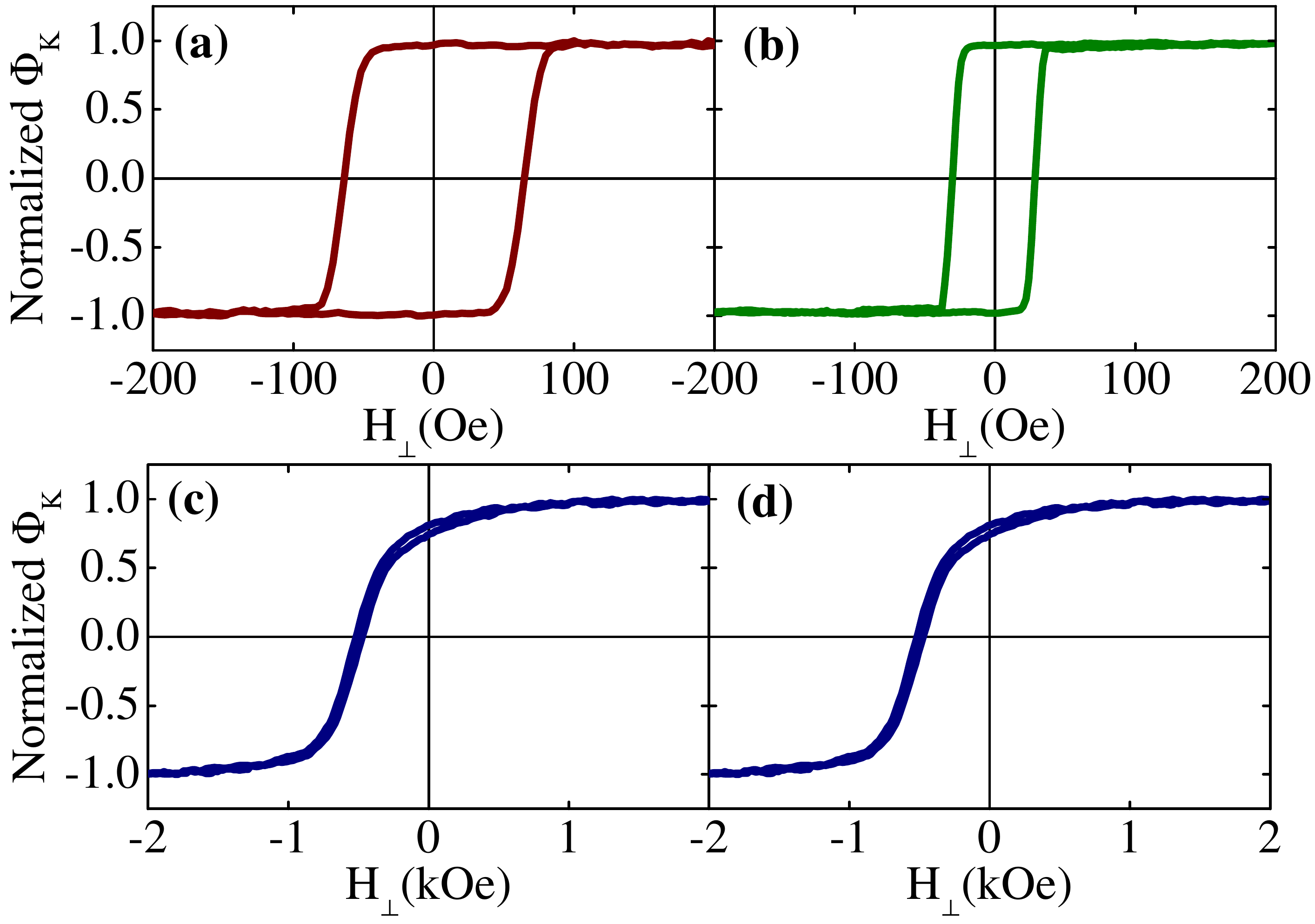}
   \caption{ Normalized hysteresis loops of the individual electrodes for the (a),(c) Hf-capped and (b),(d) Ta-capped films, acquired via MOKE at RT. Corresponding soft (a) MgO(2)/CoFeB(1.2)/Hf(5)/Pd(3) (red), (b) MgO(2)/CoFeB(1.2)/Ta(3)/Pd(3) (green) and pinned (c),(d) Ta(4)/Pd(2)/MnIr(8)/CoFe(1)/Ta(0.6)/CoFeB(0.8)/MgO(2) (blue) electrodes.}
    \label{fig:indiloop}
\end{figure}

\begin{figure}[!ht]
    \centering
    \includegraphics[height=6.7cm, width=\linewidth]{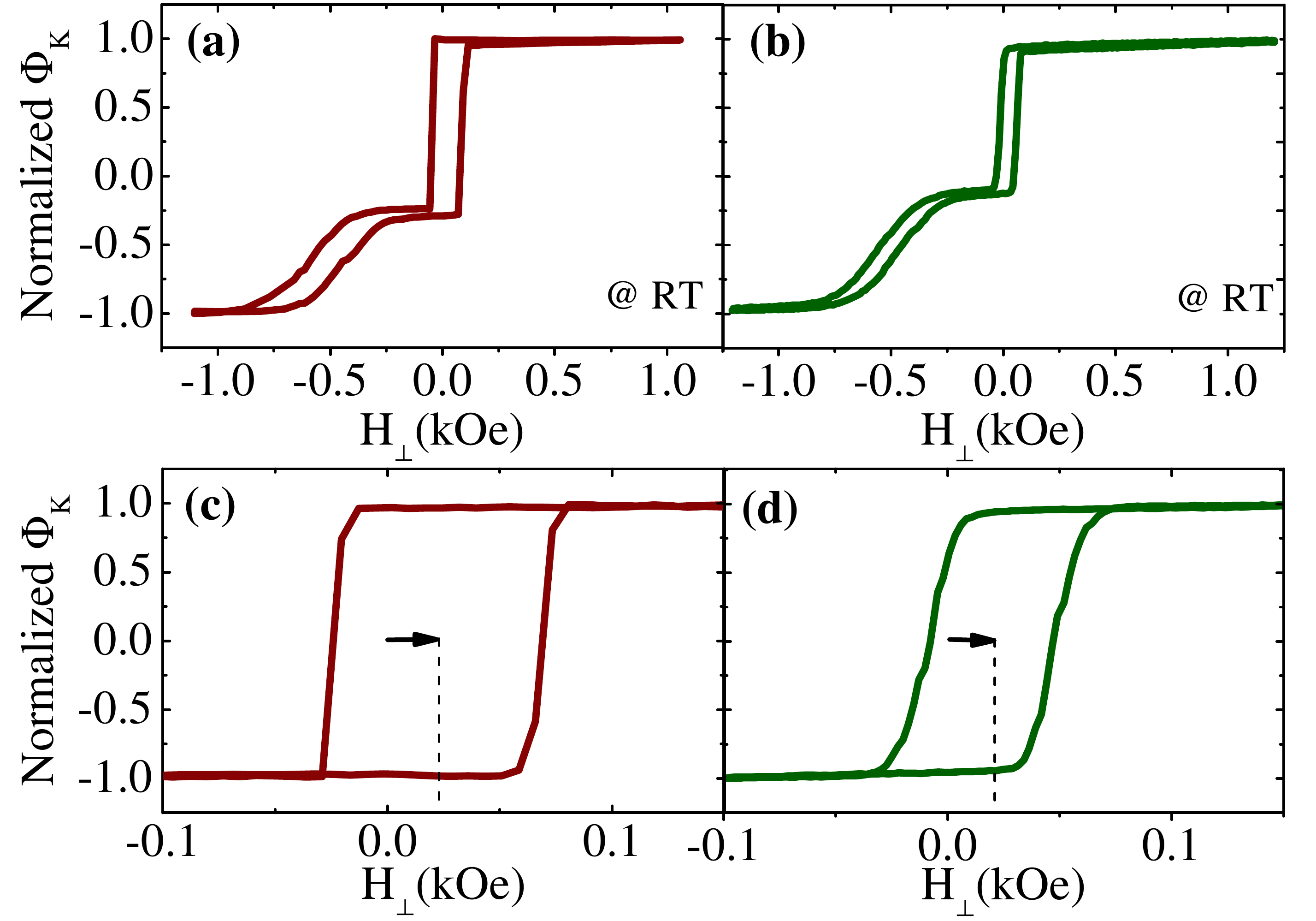}
   \caption{ Normalized (a),(b) major, (c),(d) minor perpendicular $(\perp)$ hysteresis loops of Hf (red)- and Ta (green)-capped films, respectively (MOKE at RT). }
    \label{fig:loop}
\end{figure}
\begin{figure}[!ht]
    \centering
    \includegraphics[height=5.8cm, width=\linewidth]{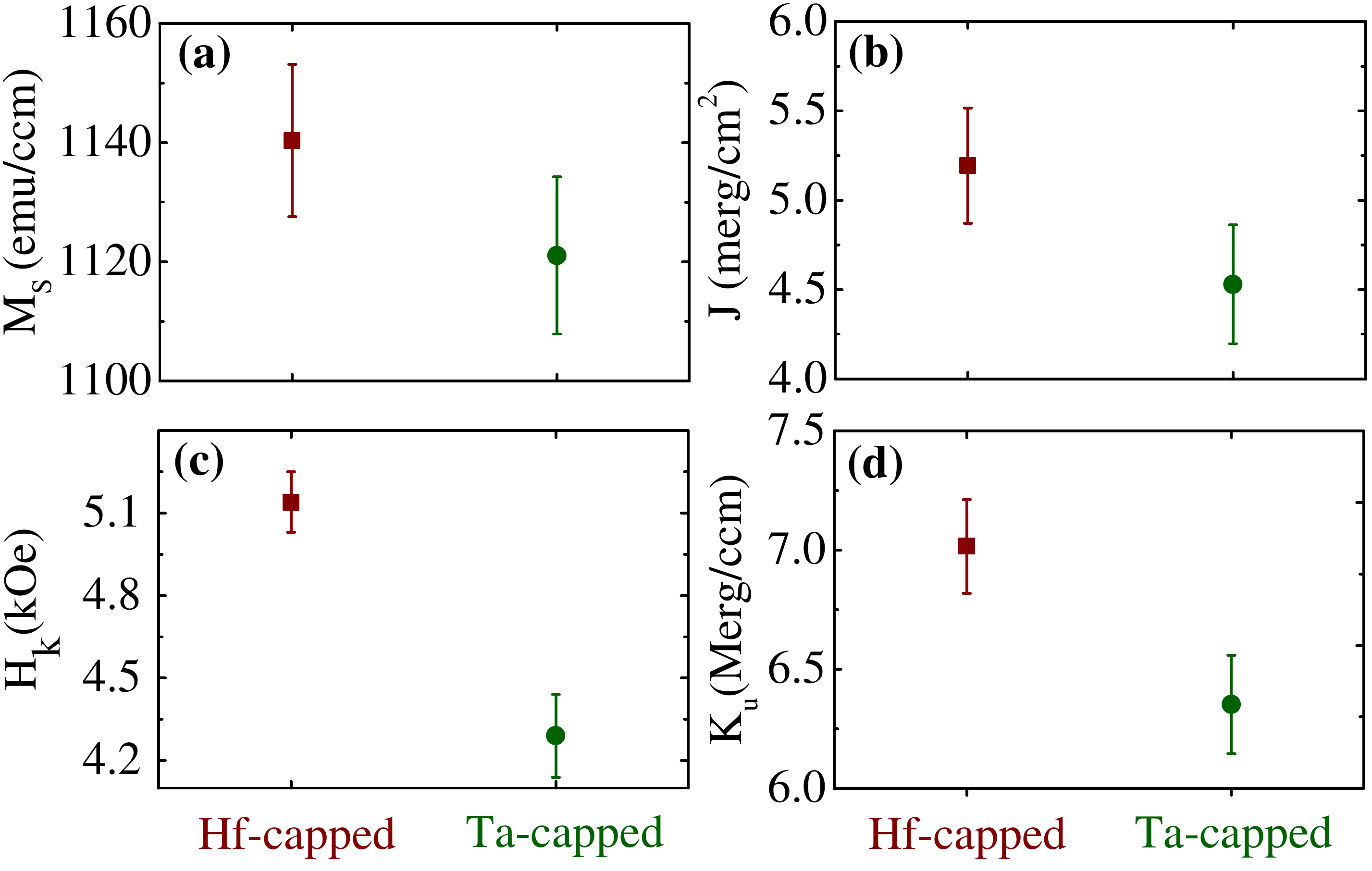}
   \caption{(a) Saturation magnetization ($M_{\textrm{s}}$). (b) Coupling constant ($J$). (c) Anisotropy field ($H_{\textrm{K}}$). (d) Uniaxial magnetic anisotropy energy $K_{\textrm{u}}$ for the Hf (red)- and Ta (green)-capped films at RT.}
    \label{fig:Magneticcharacteristics}
\end{figure}

%from the calculated saturation magnetization $M_{\textrm{s}}=(1140\,\pm\,13)\,$emu/cm$^3$$ ($M_{\textrm{s}}=(1121\,\pm\,13)\,$emu/cm$^3$), and the magnetic shift $H_{\textrm{s}}= 22$\,Oe ($H_{\textrm{s}}= 20$\,Oe), $J$ is extracted to be $J=(5.15\,\pm\,0.32)\,$merg/cm$^2$ ($J=(4.61\,\pm\,0.33)\,$merg/cm$^2$) respectively, as visible in Figs. \ref{fig:Magneticcharacteristics} (a),(b)
Figure \ref{fig:indiloop} presents the individual loops of the soft and the pinned electrodes of the Hf- and Ta-capped samples.  In particular, Figs. \ref{fig:indiloop} (a) and (b) illustrate the hysteresis loops of the soft electrodes of MgO/CoFeB/Hf/Pd (red) and MgO/CoFeB/Ta/Pd (green), respectively. Additionally, Figs. \ref{fig:indiloop} (c) and (d) show the corresponding loops of the pinned electrode Ta/Pd/MnIr/CoFe/Ta/CoFeB/MgO (blue). It is worth mentioning that the pinned part of both stacks is composed of the same sequence of materials. Furthermore, the individual films were annealed under the same conditions as the full stack, according to the description in the preparation part. From Fig. \ref{fig:indiloop} is extracted that the soft and pinned electrodes for both samples present strong PMA, while the corresponding pinned ones display an EB field of $H_{\textrm{ex}}= -500$\,Oe.\\
\mbox{ } Figure \ref{fig:loop} shows the perpendicular magnetic hysteresis loops of the total stack for the Hf (red)- and Ta (green)-capped samples, respectively. The major loops are presented in Figs. \ref{fig:loop} (a),(b) while Figs. \ref{fig:loop} (c),(d) illustrate the corresponding minor ones. In Figs.  \ref{fig:loop} (a),(b) two distinct magnetic steps are observable which correspond to the soft and the pinned electrode. As it is expected from the individual loops, the full MTJ stacks present an EB field of $H_{\textrm{ex}}= -500$\,Oe, along with PMA. It is crucial to be noted that the EB field is in a direction opposite to the applied field during annealing.\\
%The EB effect in this stack has a positive character, since the direction of the annealing field is positive ($H_{\textrm{ann.}}>0$). 
\mbox{ } Moreover, for both films (cf. Figs. \ref{fig:loop} (c),(d)) we observe a shift of the minor loops with respect to zero magnetic field. This asymmetry of the minor loop in a magnetic hysteresis measurement unveils the dipolar interactions between the soft and the pinned electrode \cite{Tsai:2014}. The coupling strength and character (ferromagnetic or antiferromagnetic) can be determined by the coupling constant $J$ which is calculated by the formula $J=\mu_\textrm{0}\,\cdot\,H_{\textrm{s}}\,\cdot\,M_{\textrm{s}}\,\cdot\,{t}$  where $\mu_\textrm{0}$ is the permeability in free space, $H_{\textrm{s}}$ is the magnetic shift of the minor loop, $M_{\textrm{s}}$ is the saturation magnetization and $t$ is the ferromagnetic thickness, respectively. It is worth noting that the calculated $M_{\textrm{s}}$ as well as the magnetic dead layer ($t_{\textrm{DL}}$) for both samples are determined from a series of films where the thickness of CoFeB in the soft electrode varies. The $M_{\textrm{s}}$ and $t_{\textrm{DL}}$ for the Hf(Ta)-capped samples are extracted to be  $M_{\textrm{s}}=(1140\,\pm\,13)\,$emu/ccm ($M_{\textrm{s}}=(1121\,\pm\,13)\,$emu/ccm) as shown in Fig. \ref{fig:Magneticcharacteristics} (a) and $t_{\textrm{DL}}=0.93\,$nm ($t_{\textrm{DL}}=0.98\,$nm), respectively. The obtained values for the $M_{\textrm{s}}$ are in good agreement with previous reports \cite{Tsai:2014}. In addition, the magnetic shift for the Hf(Ta)-capped is identified to be $H_{\textrm{s}}= 22$\,Oe ($H_{\textrm{s}}= 20$\,Oe) and consequently $J$ is extracted to be $J=(5.19\,\pm\,0.32)\,$merg/cm$^2$ ($J=(4.53\,\pm\,0.33)\,$merg/cm$^2$) respectively, as visible in Fig. \ref{fig:Magneticcharacteristics} (b). The positive value of $J$ for both samples reflects the antiferromagnetic character of coupling of both electrodes. It is already reported \cite{Weng,Moritz} that the alignment of the magnetizations of two ferromagnetic layers separated by a non-magnetic spacer prefers such type of antiferromagnetic coupling when the PMA in the system is relatively large, which promotes the magnetic volume charges (MVC) to have a dominant contribution at the determination of coupling between the two ferromagnetic layers.
In particular, a relatively strong PMA may reduce the contribution of magnetic surface charges which favor the ferromagnetic coupling while at the same time promotes the MVC which introduce strong antiparallel coupling between the ferromagnetic layers.\\
\mbox{ } A further characteristic to be pointed out is the difference between the PMA of the soft electrodes of both samples. Figs. \ref{fig:Magneticcharacteristics} (c),(d) show the anisotropy fields $H_{\textrm{K}}$ and the uniaxial magnetic anisotropy energy density $K_{\textrm{u}}$, for both samples. $H_{\textrm{K}}$ corresponds to the minimum field strength applied perpendicular to the easy axis that is able to force the magnetization to become perpendicular to the easy axis. The $K_{\textrm{u}}$ is calculated from \cite{Ikeda:2010}

\begin{equation} \label{eq:Keff}
K=K_{\mathrm{b}}-\frac{M_{\mathrm{s}}^2}{2\mu_{\mathrm{0}}}+\frac{K_{\mathrm{s}}}{t_{\mathrm{CoFeB}}}
\end{equation}
where $K$ is the perpendicular anisotropy energy density, $K_{\textrm{b}}$ is the bulk crystalline anisotropy, $K_{\textrm{s}}$ is the interfacial anisotropy and $t_\textrm{CoFeB}$ is the corresponding thickness of the CoFeB layer. The term $\frac{K_{\textrm{s}}}{t_{\textrm{CoFeB}}}$ corresponds to the $K_{\textrm{u}}$ for each sample and $K_{\textrm{b}}$ is extracted to be negligible. Consequently, the larger $K_{\textrm{u}}$ and $H_{\textrm{K}}$ values for the Hf-capped sample reflect the significantly larger PMA of the soft electrode compared to the one capped with Ta. This behaviour is in agreement with previous investigations of Hf- and Ta-capped CoFeB/MgO stacks \cite {Liu:2012} and can be attributed to the fact that Hf is a better B absorber than Ta. Furthermore, Hf promotes the crystallization of CoFe(B) in the bcc structure with (001) texture, which consequently leads to substantially higher PMA of the soft electrode.

\begin{figure}[!ht]
\centering
\includegraphics[height=6.5cm, width=\linewidth]{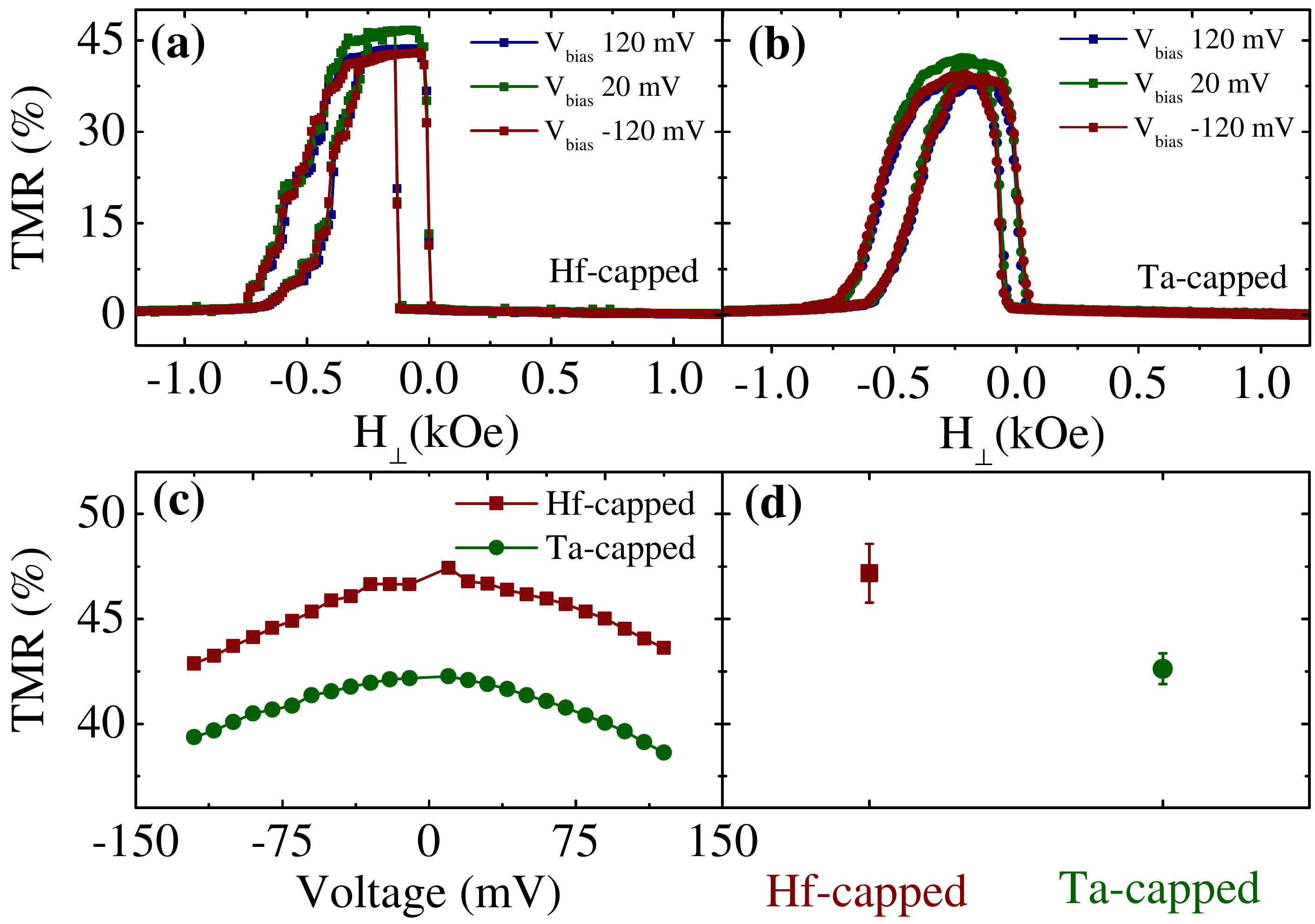}
\caption{(a),(b) Representative major TMR loops of the Hf (upper left)- and Ta (upper right)-capped samples for $V_{\textrm{bias}}=-120\,\textrm{(red)}, 20\,\textrm{(green)}, 120 \,\textrm{(blue)}$\,mV, respectively. (c) Bias dependence of TMR for Hf (red)- and Ta (green)-capped films. (d) Average TMR of six contacts acquired at $V_{\textrm{bias}}=10$\,mV for Hf (red)- and Ta (green)-capped films.}
\label{fig:TMR@RT}
\end{figure}

Figure \ref{fig:TMR@RT} summarizes the results of the TMR at RT for both samples.  
In Figs. \ref{fig:TMR@RT} (a),(b) three representative major TMR loops are displayed as a function of the perpendicular magnetic field for the Hf- and Ta-capped samples, respectively, acquired in different bias voltages ($V_{\textrm{bias}}= -120\,\textrm{(red)},\,20\,\textrm{(green)},\,120\,\textrm{(blue)}$\,mV). From similar loops at several bias voltages we calculated the TMR ratio illustrated in Fig. \ref{fig:TMR@RT} (c). After evaluating the results for six MTJ's an average TMR value is presented in Fig. \ref{fig:TMR@RT} (d) for $V_{\textrm{bias}}= 10$\,mV, for both samples. Clearly observed, the Hf-capped sample has higher TMR ratio $(47.2\pm 1.4)\%$ compared to the Ta one $(42.6\pm 0.7)\%$ at RT. This is consistent with the claim of J. D. Burton $et\,al.$ \cite {Burton:2006} that the presence of B at the CoFeB/MgO interface, suppresses the coherent tunneling in the $\Delta_{\textrm{1}}$ band, leading to the reduction of TMR. Thus, preventing the presence of B at the interface should enhance the TMR in these junctions. Moreover, this is in agreement with the fact that Hf is a better B absorber material than Ta, which can be, e.g., concluded from the calculated values of metal Boride enthalpies \cite {Hindmarch:2011}. The predicted formation enthalpies \cite{Niessen} of Hf (Ta) borides that may be anticipated within a typical MTJ are $\Delta H_{\textrm{HfB}}\,=\,-95$\,kJ/mol, ($\Delta H_{\textrm{TaB}}\,=\,-78$\,kJ/mol), $\Delta H_{\textrm{Hf}_{2}\textrm{B}}\,=\,-67$\,kJ/mol, ($\Delta H_{\textrm{Ta}_{2}\textrm{B}}\,=\,-56$\,kJ/mol), $\Delta H_{\textrm{HfB}_{2}}\,=\,-95$\,kJ/mol, ($\Delta H_{\textrm{TaB}_{2}}\,=\,-83$\,kJ/mol), respectively. This underpins that Hf will lead to a stronger absorption of B than Ta.

%In Fig. \ref{fig:RTcomparison} (a) a comparison of TMR is illustrated for Hf and Ta capped samples. It is clear that the Hf capped sample has higher TMR ratio compared to the Ta. This is consistent with the claim of J. D. Burton $et\,al.$  \cite {Burton:2006}  that the presence of B at the CoFeB/MgO interface, suppresses the coherent tunneling in the $\Delta$1 band leading to the reduction of TMR. Thus, preventing the presence of B at the
%interface should enhance the TMR in these junctions.
%Moreover, this is in agreement with the fact that Hf is a better B absorber material than Ta, which can be also concluded from the calculated values of metal Boride enthalpies \cite {Hindmarch:2011}. The predicted formation enthalpies \cite{Niessen} of Hf and (Ta) borides which may be anticipated within a typical MTJ are $\Delta H_{\textrm{HfB}} =-95\,kJ/mol$, ($\Delta H_{\textrm{TaB}} = -78\,kJ/mol)$, $\Delta H_{\textrm{Hf}_{2}\textrm{B}}=-67\,kJ/mol$, ($\Delta H_{\textrm{Ta}_{2}\textrm{B}}=-56\,kJ/mol)$, $\Delta H_{\textrm{HfB}_{2}}=-95\,kJ/mol$, ($\Delta H_{\textrm{TaB}_{2}}=-83\,kJ/mol)$ respectively. This underpins the opinion that Hf will lead to a larger absoption of B than Ta. In Fig. \ref{fig:RTcomparison} (b) (\ref{fig:RTcomparison} (c)) three representative major loops are displayed for different bias voltages for the Hf (Ta) capped samples, respectively. From these loops at different bias voltages we calculated the TMR ratio presented on Fig. \ref{fig:TMRvsT} (a).

\begin{figure}[!htb]
    \centering
    \includegraphics[height=6.5cm, width=\linewidth]{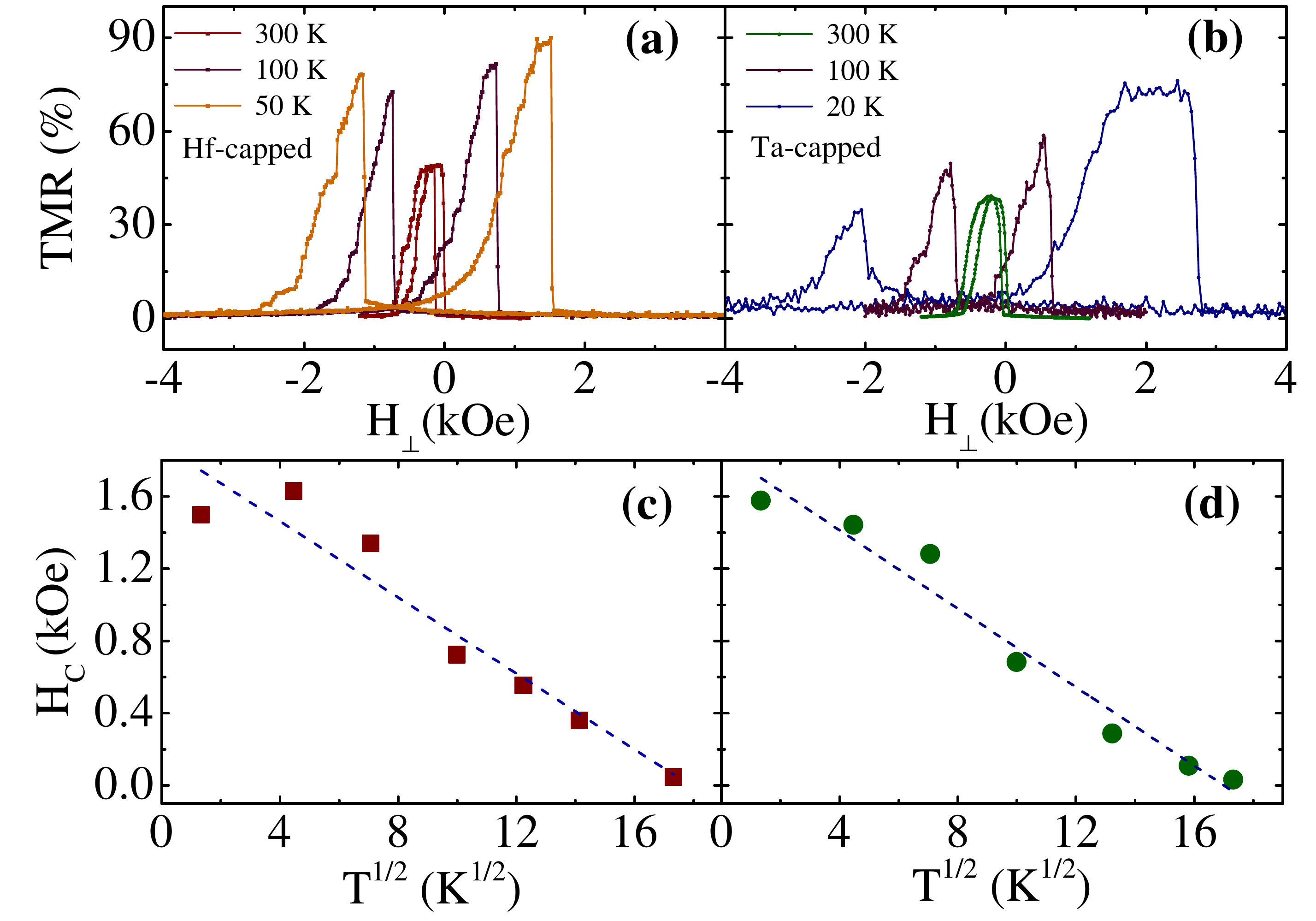}
   \caption{(a),(b) Major TMR loops of the Hf (upper left)- and (Ta) (upper right)-capped samples for $V_{\textrm{bias}}=20\,(60)$\,mV at $T=50\,(20), 100\,(100), 300\,(300)$\,K, respectively. (c), (d) $H_{\textrm{c}}$ of the soft electrode versus $T^{1/2}$ (squares: experimental values, dashed line: model following Eq. (\ref{eq:coercivities}) for the Hf (red)- and Ta (green)-capped films, respectively.}
    \label{fig:TMRvsT}
\end{figure}

Figures \ref{fig:TMRvsT} (a),(b) present the dependencies of the TMR
on the external perpendicular field for Hf (Ta)-capped samples at different temperatures $T=50\,(20), 100\,(100), 300\,(300)$\,K for $V_{\textrm{bias}}=20\,(60)$\,mV, respectively. In Figs. \ref{fig:TMRvsT} (c),(d) the $H_{\textrm{c}}$ of the soft electrodes of the Hf- and Ta-capped samples, which were extracted from the corresponding minor TMR loops (not shown) are plotted as a function of $T^{1/2}$ . The temperature dependent behavior of $H_{\textrm{c}}$ for both samples can be described by Stoner-Wohlfarth (SW) model \cite{stoner} under thermal fluctuations. In this model the temperature dependence of $H_{\textrm{c}}$ is given by \cite{Nunes}
\vspace{-0.4em}

\begin{equation} \label{eq:coercivities}
H_{\textrm{c}}=H_{\textrm{c0}}[1-({\frac{T}{T_{\textrm{B}}}})^{1/2}]
\end{equation}
where $T_{\textrm{B}}$ is the blocking temperature and $H_{\textrm{c0}}$ is the coercivity at 0\,K. The extracted fitting parameters for the Hf (Ta)-capped samples are: $H_{\textrm{c0}}=(1.88\,\pm\,0.14)$\,kOe ($H_{\textrm{c0}}=(1.84\,\pm\,0.10)$\,kOe) and $T_{\textrm{B}}=318.4$\,K ($T_{\textrm{B}}=289.2$\,K), respectively.
For both samples, the experimentally observed values for $H_{\textrm{c}}$ are in reasonable agreement with the values predicted by Eq. (\ref{eq:coercivities}). However, some slight deviations are observed especially at low temperatures. One reason could be the interaction of the soft electrode with the reference system that is also temperature
dependent and prefers the antiparallel state, thereby adding an extra torque to the soft layers' magnetization. Another option is a magnetization reversal via
domain wall nucleation and movement, that could induce an exponential dependence of $H_{\textrm{c}}$ on T. If only one or more mechanisms are responsible for the experimental results will be investigated in further experiments.

%\ref{eq:name}

%The \ref{fig:TMRvsT} and \ref{fig:TMRTavsT} present the dependencies of the TMR on the external perpendicular field (a), on the bias voltage (b) and on the temperature (c) as well as the dependence of the coercive field Hc on $T^{1/2}$ (d) for Hf- \ref{fig:TMRHfvsT} and Ta-capped \ref{fig:TMRTavsT} samples. The Fig. \ref{fig:TMRHfvsT} (b) (\ref{fig:TMRTavsT} (b)) and \ref{fig:TMRHfvsT} (c) ((\ref{fig:TMRTavsT}(c)) show the commonly observed increase of the TMR with decreasing temperature. The largest observed TMR values are 105$\%$ (125$\%$) at 10\,K\,(1.8\,K) and  $V_{\textrm{bias}}=10\,(20)\,mV$ for the Hf (Ta) capped sample. A further feature which can be extracted from the graph is that at $T=300\,K$ the Hf capped sample presents a larger TMR value of $(51.4\pm 2.1)\%$ compared to the Ta one $(43.9\pm 3.8)\%$. 
%This result has been determined by a large number of characterized contacts and supports the suggestion that the improved boron absorption of Hf causes the increase of the TMR. Additionally, it can be commented that there is a trend of higher TMR values for the Hf capped sample compared to the Ta one at the temperature range $T=10-300\,K$. However, this claim is not statistically strong enough due to the small number of characterized contacts.%shown in Fig. \ref{fig:comparisonTMR}.
\par

%The temperature and voltage dependencies of the TMR can be interpreted by several models. The model of bias dependent magnon assisted tunneling \cite{Zhang:1997} suggests that at the ferromagnet/insulator interface inelastic electron scattering by magnon excitations take place that modulates the voltage dependence of the TMR. Phenomenologically, this can be described as follows. Electrons, which have tunneled across the barrier, under the presence of non-zero bias, arrive at the second ferromagnet as hot electrons with energies higher than the Fermi energy of this electrode (provided that no inelastic scattering event has occurred inside the tunnel barrier). Then these hot electrons lose their energy by emitting a magnon and thereby flipping their spin. With increasing bias voltage more magnons can be emitted resulting in reduced TMR values. 
\par

\vspace{-0.7em}

\section{Conclusion}
In summary, we investigated the magneto-transport properties of p-MTJs with PEB stacks Ta/Pd/IrMn/CoFe/Ta/CoFeB/MgO/CoFeB/Hf/Pd and Ta/Pd/IrMn/CoFe/Ta/CoFeB/MgO/CoFeB/Ta/Pd. 
The Hf- and Ta-capped stacks showed a PEB of $H_\textrm{ex}=-500$\,Oe along with PMA having TMR values of $(47.2\pm 1.4)\%$ and $(42.6\pm 0.7)\%$ at RT, respectively. The larger PMA and TMR values, for the Hf- compared to the Ta-capped sample were attributed to the enhanced B absorption of Hf. Additionally, the temperature dependence of the $H_{\textrm{c}}$ of the soft electrodes was described by the Stoner-Wolfram model while the observed slight deviation from the model for both samples was interpreted qualitatively by an additional torque from the interactions occurring between the AF/FM double layer and the soft electrode. 

\vspace{-0.7em}

\section{Acknowledgment}

The authors would like to thank for the financial support HARFIR and the Deutsche Forschungsgemeinschaft (DFG, contract RE1052/32).

\bibliographystyle{apsrev4-1}
\bibliography{main}

\end{document}